\documentclass{pasa}%

\usepackage{graphicx}

\title[Radio continuum from massive galaxies]{Radio continuum from the most massive early-type galaxies detected with ASKAP RACS}

\author[Brown et al.]{Michael J. I. Brown$^1$, Teagan A. Clarke$^{1,2}$, Andrew M. Hopkins$^3$, Ray P. Norris$^{4,5}$ and T.H.~Jarrett$^{6}$ 
\affil{$^1$ School of Physics \& Astronomy, Monash University, Clayton, VIC 3800, Australia}%
\affil{$^2$ OzGrav: The ARC Centre of Excellence for Gravitational-wave Discovery, Clayton, VIC 3800, Australia}
\affil{$^3$ School of Mathematical and Physical Sciences, 12 Wally’s Walk,
Macquarie University, NSW 2109, Australia}
\affil{$^4$ School of Science, Western Sydney University, Locked Bag 1797, Penrith South DC, NSW 1797, Australia}
\affil{$^5$ CSIRO Space \& Astronomy, PO Box 76, Epping, NSW 1710, Australia}
\affil{$^6$ Astronomy Dept, University of Cape Town, Rondebosch 7701, Western Cape, South Africa}

}%

\jid{PASA}
\doi{10.1017/pas.\the\year.xxx}
\jyear{\the\year}

\usepackage{aas_macros}
\usepackage{hyperref} 
\hypersetup{colorlinks,citecolor=blue,linkcolor=blue,urlcolor=blue}

\hypersetup{draft}

\begin{document}

\begin{frontmatter}
\maketitle

\begin{abstract}
All very massive early-type galaxies contain supermassive blackholes but are these blackholes all sufficiently active to produce detectable radio continuum sources? We have used the 887.5~MHz Rapid ASKAP Continuum Survey DR1 to measure the radio emission from morphological early-type galaxies brighter than $K_S=9.5$ selected from the 2MASS Redshift Survey, HyperLEDA and RC3. In line with previous studies, we find median radio power increases with infrared luminosity, with $P_{1.4} \propto L_K^{2.2}$, although the scatter about this relation spans several orders of magnitude. All 40 of the $M_K<-25.7$ early-type galaxies in our sample have measured radio flux densities that are more than $2\sigma$ above the background noise, with $1.4~{\rm GHz}$ radio powers spanning $\sim 3 \times 10^{20}$ to $\sim 3\times 10^{25}~{\rm W~Hz^{-1}}$. Cross matching our sample with integral field spectroscopy of early-type galaxies reveals that the most powerful radio sources preferentially reside in galaxies with relatively low angular momentum (i.e. slow rotators). While the infrared colours of most galaxies in our early-type sample are consistent with passive galaxies with negligible star formation and the radio emission produced by active galactic nuclei or AGN remnants, very low levels of star formation could power the weakest radio sources with little effect on many other star formation rate tracers. 
\end{abstract}
\end{frontmatter}

\section{INTRODUCTION}
\label{sec:intro}


The presence of supermassive blackholes (SMBHs) with masses of millions or billions of Solar masses, in all very massive galaxies has been established for over two decades \citep{magorrian1998}. While there are many well-known examples of radio-loud active galactic nuclei (AGNs) in these galaxies, including M~87 and Fornax~A, it is unclear if all SMBHs in very massive galaxies are undergoing sufficient accretion to have corresponding detectable radio continuum emission.  


Correlations between AGN radio power and host galaxy mass have been measured (or implied) by numerous studies for at least three decades. The K-z relation for powerful radio sources, where $K$-band apparent magnitude is correlated with redshift \citep{lilly1984}, is a consequence of the most powerful radio-loud AGN hosts being limited to the top end of the galaxy stellar mass function \citep[e.g.][]{rocca2004}. While the K-z relation has relatively little dispersion in shallow radio surveys such as 3C, deeper radio surveys show more dispersion as less powerful radio sources are hosted by galaxies with a broader range of stellar masses \citep[e.g.][]{rocca2004}.


The radio powers of early-type galaxies as a function of absolute magnitude and mass has been measured at low redshift by a number of studies since the late-1980s \citep[e.g.][]{fabbiano1989, sadler1989, best2005, best2007, brown2011, sabater2019, capetti2022}. \citet{fabbiano1989} and \citet{sadler1989} used targeted radio observations of nearby early-type galaxies to measure the strong correlation between radio power and $B$-band luminosity for early-type galaxies, and demonstrated that the distribution of radio powers at fixed $B$-band luminosity spans at least four orders of magnitude. However, while \citet{fabbiano1989} and \citet{sadler1989} could detect radio sources in many of the most massive galaxies, they were unable to detect radio sources in all of these galaxies. 


Should the broad distribution of radio powers be interpreted as a duty cycle, with the radio powers of individual objects varying by orders of magnitudes, or do individual objects retain high or low radio powers for billions of years? If the former is the case, then we would expect the high and low power radio sources to reside in host galaxies and groups with comparable properties, including morphology, stellar kinematics and dark matter halo masses (i.e. a highly variable radio source will not change its host galaxy's mass, shape and large-scale environment). Conversely, if the latter is the case then high and low power radio sources will reside in galaxies with significantly different properties (i.e. morphology, kinematics or environment) at fixed stellar mass.

S0 galaxies typically have lower radio powers than comparable mass elliptical galaxies \citep[e.g.][]{hummel1982,veron2001}, but even within the elliptical class galaxies with boxy or round isophotes typically have higher radio powers than galaxies with pointed isophotes or disk components \citep[e.g.][]{hummel1983,bender1987,bender1989}. \citet{zheng2023} recently found that the kinematic properties of galaxies with high and low radio powers (at fixed stellar mass) differ from each other, with the most powerful radio sources being preferentially hosted by galaxies with low angular momentum (as measured by integral field spectroscopy). There are a number of proxies for dark matter halo mass and constraints on models of dark matter halo mass, including large-scale environment (measured with galaxy clustering), gravitational lensing and X-ray luminosity \citep[e.g.][and references therein]{benson2000}. However, there are conflicting conclusions in the literature about whether high power radio sources have dark matter halo masses that are the same as \citep[e.g.][]{magliocchetti2004,worpel2013} or higher than \citep[e.g.][]{mandelbaum2009} other galaxies with the same stellar mass. 


Wide-field radio continuum surveys including NRAO VLA Sky Survey (NVSS), Faint Images of the Radio Sky at Twenty centimeters survey (FIRST) and LOFAR Two-metre Sky Survey (LoTSS) have enabled measurements of the relationship between AGN radio power and host galaxy mass using low redshift galaxy catalogues and surveys, including Third Reference Catalogue of Bright Galaxies (RC3) and the Sloan Digital Sky Survey \citep[e.g.][]{best2005,best2007,brown2011,gurkan2018,sabater2019,capetti2022}. Both \citet{brown2011} and \citet{sabater2019} find that the measured radio flux densities of very massive early-type and passive galaxies are {\it consistent} with them being all being radio sources, but the conclusions of these studies are limited by the depth of their radio data and sample size respectively. 


Recently \citet{grossova2022} and \citet{capetti2022} have detected large fractions of nearby early-type galaxies with the Karl G. Jansky Very Large Array (VLA) and the Low Frequency Array (LOFAR). \citet{grossova2022} obtained 1-2 GHz detections of 41 of 42 nearby $M_B<-20$ early-type galaxies associated with $L_X>10^{40}~{\rm erg~s^{-1}}$ X-ray sources, with the one remaining galaxy (NGC~499) having a 150~MHz detection with LOFAR. \citet{capetti2022} used LoTSS to obtain 150~MHz radio detections of 146 (78\%) of 188 $M_K<-25$ early-type galaxies, and detect all 25 early-type galaxies in their sample with $M_K<-25.8$. \citet{capetti2022} caution that while they have a 100\% detection rate for the most massive galaxies in their sample, four of these detections are associated with remnants of active galactic nuclei and at least one is likely to be dominated by star formation.


To determine if all very massive early-type galaxies are radio continuum sources, the obvious approach is to combine (relatively) deep radio imaging with large samples of nearby early-type galaxies \citep[e.g.][]{brown2011, capetti2022, grossova2022}. The Rapid ASKAP Continuum Survey \citep[RACS; ][]{racs2020, hale2021, duchesne2023} is a continuum survey of the sky south of ${\rm Decl.}$ $+47^\circ$ in three frequency bands (RACS-low: 887.5 MHz, RACS-mid: 1367.5 MHz and RACS-high: 1655.5 MHz). For this work we are using RACS-low DR1 \citep{hale2021}, which provides 887.5 MHz images of the sky in the declination range  $-84.1^\circ \delta < +30.0^\circ$ convolved to a common resolution of $25^{\prime\prime}$ with a typical RMS depth $300~{\rm \mu Jy/beam}$.  Given radio continuum spectra are often approximated $f_\nu \propto \nu^{-0.7}$ and the NVSS depth is typically $450~{\rm \mu Jy/beam}$, RACS is effectively a factor of $\sim 2$ deeper than the $1.4~{\rm GHz}$ NVSS. The 2MASS Redshift Survey \citep[2MRS;][]{huchra2012} provides redshifts for nearly all $K_S\leq 11.25$ galaxies with $|b|>10^\circ$, and is thus the ideal parent sample for selecting nearby galaxies. For bright nearby galaxies reliable galaxy morphologies are available via the Third Reference Catalogue of Bright Galaxies \citep[RC3 ;][]{rc3} and HyperLEDA \citep{paturel2003}.


In this work we use RACS, 2MRS, RC3 and HyperLEDA to determine if all very massive nearby early-type galaxies are radio sources. In Section~\ref{sec:data} we describe the data, sample selection and radio flux density measurements. We present our results in Section~\ref{sec:results}, including measurements of radio continuum flux density and radio power as a function of $K$-band absolute magnitude. We discuss how our results compare to the literature in Section~\ref{sec:discussion}, including the fraction of early-type galaxies that are radio sources as a function of K-band absolute magnitude and stellar mass. Finally, our conclusions are provided in Section~\ref{sec:conclusion}. Throughout this paper we use Vega magnitudes, approximate early-type radio continuum with $f_\nu \propto \nu^{-0.7}$ and use the bulk flow model of \citet{carrick2015}.
  
\section{Data and measurements}
\label{sec:data}


Our galaxy sample is drawn from the subset of 2MRS in the declination range  $-84.1^\circ < \delta < +30.0^\circ$, corresponding to the region of 2MRS that could overlap with the RACS DR1 images or catalogues of \citet{hale2021}. The full 2MRS provides redshifts and 2MASS photometry for 97.6\% of galaxies brighter than $K_S=11.75~{\rm mag}$ over a region encompassing 91\% of the sky. While 2MRS does provide galaxy morphologies, we found some Sa galaxies were misclassified as early-type galaxies, so we use HyperLEDA and RC3 morphologies throughout the remainder of the paper. 

To obtain a sample with a high fraction of radio detections, we select a bright early-type subset of 2MRS with $K_S<9.5 ~{\rm mag}$ and 2003 HyperLEDA \citep{paturel2003} classifications of E, E-SO or SO. For a small minority of galaxies where 2003 HyperLEDA classifications were unavailable we selected galaxies if they had RC3 classifications of E or L (Elliptical or Lenticular). We checked the sample for contamination by spiral galaxies using SDSS, PanSTARRS or Skymapper images \citep{york2000,chambers2016,wolf2018} and after removing obvious contaminants we selected 640 early-type galaxies with $-84.1^\circ \delta < +30.0^\circ$. Our early-type galaxies are sufficiently nearby (median distance of 41 Mpc) that bulk flows can be comparable to the Hubble flow, so we determine distances using the 2MRS redshifts and the bulk flow corrections of \citet{carrick2015}\footnote{https://cosmicflows.iap.fr} or, when available, redshift independent distances from Cosmicflows-3 \citep{tully2016}



For each galaxy in our sample we measure the flux density per beam and (if needed) the integrated flux density from RACS images cutouts. The cutouts were extracted using astropy \citep{astropy2013, astropy2018}, astroquery and Jupyter notebooks that build on code provided by CSIRO ASKAP Science Data Archive (CASDA). The RACS image cutouts are typically $15^{\prime} \times 15^{\prime}$ for $K_S<7$ galaxies and $5^{\prime} \times 5^{\prime}$ for $K_S>7$ galaxies, but all cutouts have been visually inspected and larger cutouts were extracted when needed. Of the 640 early-type galaxies in the parent sample, 633 have at least one ASKAP RACS DR1 cutout image.



For each galaxy the flux density per beam is measured directly from the pixel corresponding to the galaxy position, while for extended sources an elliptical aperture is used to measure the integrated flux density. The elliptical aperture is defined manually via visual inspection of the RACS images, to account for multi-component sources. We manually removed NGC 4783 from the sample as its measured RACS flux density is corrupted by neighbouring NGC 4782. To quantify noise and its impact we also measure the median absolute deviation (MAD) for each image cutout and the flux density per beam at a position offset by 25 pixels ($\simeq 1^\prime$) in both image axes from each galaxy. As a small portion of RACS images have relatively high background noise, for the remainder of the analysis we restrict ourselves to the 587 early-type galaxies that have a $>2\sigma$ flux density per beam detection or RACS cutouts with a median absolute deviation less than $500~{\rm \mu Jy}/{\rm beam}$. 



As interferometers such as ASKAP can underestimate the flux densities of bright and very extended radio sources, we also cross match our sample against the Molonglo Observatory Synthesis Telescope (MOST) sample of \citet{jones1992} and single dish data from Parkes \citep{wright1990} and the Green Bank 300-ft \citep{white1992}. We use single dish data when it provides a flux density greater than 1~Jy and the measured RACS flux density is greater than 0.1~Jy (while assuming $f_\nu \propto \nu^{-0.7}$ to account for frequency differences). Our galaxy sample has significant overlap with the $K<9$ early-type galaxy sample of \citet{brown2011}, which makes use of the NVSS and single dish data. While RACS is deeper than NVSS, the two surveys should be broadly consistent and we thus use the \citet{brown2011} sample to flag potentially erroneous measurements.


\section{Results}
\label{sec:results}

\begin{figure}
\begin{center}
\includegraphics[width=\columnwidth]{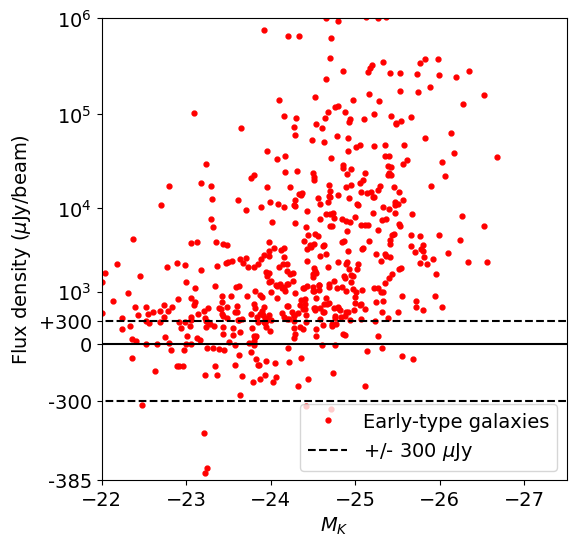}
\caption{Flux density per beam of $K_S<9.5$ early-type galaxies as a function of $K_S$-band absolute magnitude. The dashed lines show the approximate $1\sigma$ noise of the RACS images, and 408 of the 587 early-type galaxies in our sample are detected with $>2\sigma$ significance. The absence of very massive early-type galaxies with negative flux densities is consistent with all very massive early-type galaxies being radio sources.}\label{fig:fnu}
\end{center}
\end{figure}

The flux densities our 587 $K_S<9.5$ early-type galaxies as a function of absolute magnitudes are presented in Figure~\ref{fig:fnu}. Key measurements for individual galaxies (ordered by apparent $K_S$ magnitude) are provided in Table~\ref{table:RACSellipticals} including RACS flux density per beam, RACS median absolute deviation (noise), RACS integrated flux density, literature flux density (where relevant) and distance. 
We find 408 (70\%) of the 587 sample galaxies have a flux density measurement with greater than $2\sigma$ significance. A strong trend with absolute magnitude can be seen, with some of the most massive galaxies in the sample having continuum flux densities of $>0.1~{\rm Jy}$. While some of the least massive galaxies in the sample have negative measured flux densities (due to noise), the most massive galaxies in the sample  have positive radio continuum flux densities. All 40 $M_K<-25.7$  early-type galaxies in our sample have RACS detections with more than $2\sigma$ significance.


\begin{table*} 
\caption{The RACS sample of nearby elliptical galaxies (full table is available online).}
\label{table:RACSellipticals}
\scriptsize
\begin{tabular}{cccccccccccc}
\hline
 Name & R.A.     & Decl.   &  RACS  & RACS   & RACS   & Lit. & Source & $K_S$ &         $d$ & ${\rm log} P_{1.4}$ & $M_K$ \\ 
      & (J2000)  & (J2000)   &  $f_{0.8875, {\rm beam}}$ & MAD   & $f_{0.8875, {\rm int}}$  & $f_{0.8875, {\rm int}}$ & & (mag) &         (Mpc)& (W/Hz) & (mag)\\ 
     &  &  &  (mJy)                 & (mJy) & (mJy)                 & (mJy)   &  &  &  &           &  \\        
    \hline
NGC5128                 &  201.3656 & -43.0187 &   5339 &   13.5 &  $   142 \times 10^3 $ &  $   378 \times 10^3 $ &  MOST       &   3.95 &   3.66 &  23.64 &  -23.87 \\ 
M49                     &  187.4450 &   8.0004 &    172 &   0.23 &    335 &    -   &  RACS       &   5.50 &  16.07 &  21.88 &  -25.53 \\ 
NGC1316                 &   50.6741 & -37.2082 &    268 &   0.23 &   3531 &  $   163 \times 10^3 $ &  MOST       &   5.68 &  17.46 &  24.64 &  -25.53 \\ 
M60                     &  190.9167 &  11.5526 &   22.8 &   1.01 &    -   &    -   &  RACS       &   5.82 &  17.38 &  20.78 &  -25.38 \\ 
M87                     &  187.7059 &  12.3911 &  51780 &   14.0 &  $   222 \times 10^3 $ &  $   304 \times 10^3 $ &  Parkes     &   5.90 &  16.52 &  24.86 &  -25.19 \\ 
NGC3115                 &  151.3082 &  -7.7186 &   1.12 &   0.18 &   1.26 &    -   &  RACS       &   5.92 &  10.28 &  19.06 &  -24.14 \\ 
M85                     &  186.3502 &  18.1911 &   0.54 &   0.20 &    -   &    -   &  RACS       &   6.25 &  15.85 &  19.07 &  -24.75 \\ 
NGC2974                 &  145.6387 &  -3.6991 &   11.7 &   0.22 &    -   &    -   &  RACS       &   6.25 &  22.08 &  20.70 &  -25.47 \\ 
NGC 2784                &  138.0812 & -24.1726 &   1.18 &   0.21 &    -   &    -   &  RACS       &   6.32 &   9.59 &  18.97 &  -23.59 \\ 
M84                     &  186.2657 &  12.8870 &    940 &   1.71 &   8251 &   8933 &  300-ft     &   6.33 &  16.83 &  23.34 &  -24.80 \\ 
NGC 1553                &   64.0436 & -55.7801 &   3.84 &   0.15 &    -   &    -   &  RACS       &   6.34 &  15.63 &  19.91 &  -24.63 \\ 
M105                    &  161.9567 &  12.5816 &   0.83 &   0.23 &    -   &    -   &  RACS       &   6.35 &  11.22 &  18.96 &  -23.90 \\ 
NGC1399                 &   54.6212 & -35.4507 &    138 &   0.46 &    720 &    -   &  RACS       &   6.44 &  22.08 &  22.48 &  -25.28 \\ 
NGC4697                 &  192.1496 &  -5.8008 &   1.39 &   0.35 &    -   &    -   &  RACS       &   6.49 &  12.42 &  19.27 &  -23.98 \\ 
NGC4526                 &  188.5126 &   7.6991 &   14.4 &   0.21 &   17.5 &    -   &  RACS       &   6.54 &  15.00 &  20.53 &  -24.34 \\ 
NGC 3923                &  177.7574 & -28.8062 &   1.69 &   0.17 &    -   &    -   &  RACS       &   6.61 &  20.51 &  19.79 &  -24.95 \\ 
NGC4636                 &  190.7078 &   2.6878 &   73.1 &   0.40 &    108 &    -   &  RACS       &   6.62 &  15.07 &  21.33 &  -24.27 \\ 
NGC2640                 &  129.3526 & -55.1237 &   29.9 &   0.17 &   34.0 &    -   &  RACS       &   6.78 &  10.06 &  20.48 &  -23.23 \\ 
NGC 3585                &  168.3213 & -26.7550 &   0.56 &   0.18 &    -   &    -   &  RACS       &   6.79 &  18.79 &  19.23 &  -24.58 \\ 
NGC4365                 &  186.1176 &   7.3175 &   0.83 &   0.18 &    -   &    -   &  RACS       &   6.79 &  22.80 &  19.58 &  -25.00 \\ 
NGC2663                 &  131.2844 & -33.7948 &    270 &   0.31 &   2315 &   3041 &  Parkes     &   6.79 &  27.60 &  23.30 &  -25.41 \\ 
NGC1407                 &   55.0496 & -18.5804 &   95.9 &   0.58 &    123 &    -   &  RACS       &   6.83 &  28.18 &  21.93 &  -25.42 \\ 
M89                     &  188.9162 &  12.5560 &   95.7 &   1.63 &    -   &    -   &  RACS       &   6.84 &  15.85 &  21.32 &  -24.16 \\ 
NGC3384                 &  162.0704 &  12.6293 &   0.03 &   0.22 &    -   &    -   &  RACS       &   6.84 &  10.05 &  17.47 &  -23.17 \\ 
NGC 1549                &   63.9384 & -55.5924 &   0.45 &   0.14 &    -   &    -   &  RACS       &   6.88 &  17.38 &  19.07 &  -24.32 \\ 
NGC4976                 &  197.1564 & -49.5064 &   1.19 &   0.30 &    -   &    -   &  RACS       &   6.88 &  12.61 &  19.22 &  -23.62 \\ 
NGC 1404                &   54.7163 & -35.5944 &   3.92 &   0.42 &   4.63 &    -   &  RACS       &   6.90 &  18.79 &  20.15 &  -24.47 \\ 
IC 1459                 &  344.2945 & -36.4625 &   1042 &   0.34 &   1088 &    -   &  RACS       &   6.92 &  28.71 &  22.89 &  -25.37 \\ 
NGC 1380                &   54.1150 & -34.9763 &   4.30 &   0.28 &    -   &    -   &  RACS       &   6.96 &  18.62 &  20.11 &  -24.39 \\ 
NGC 1395                &   54.6241 & -23.0277 &   4.34 &   0.15 &    -   &    -   &  RACS       &   7.02 &  23.77 &  20.33 &  -24.86 \\ 
NGC 5846                &  226.6221 &   1.6054 &   23.1 &   0.24 &    -   &    -   &  RACS       &   7.08 &  25.00 &  21.10 &  -24.91 \\ 
NGC3607                 &  169.2277 &  18.0518 &   8.75 &   0.20 &    -   &    -   &  RACS       &   7.11 &  22.49 &  20.59 &  -24.65 \\ 
NGC 6684                &  282.2412 & -65.1734 &   0.09 &   0.14 &    -   &    -   &  RACS       &   7.11 &  13.55 &  18.15 &  -23.55 \\ 
NGC 1332                &   51.5722 & -21.3354 &   4.20 &   0.38 &    -   &    -   &  RACS       &   7.11 &  24.55 &  20.34 &  -24.84 \\ 
NGC4494                 &  187.8501 &  25.7750 &   0.98 &   0.22 &    -   &    -   &  RACS       &   7.14 &  16.90 &  19.39 &  -24.00 \\ 
NGC5102                 &  200.4902 & -36.6302 &   1.28 &   0.19 &    -   &    -   &  RACS       &   7.18 &   3.73 &  18.19 &  -20.68 \\ 
NGC 5084                &  200.0705 & -21.8276 &   33.3 &   0.22 &    -   &    -   &  RACS       &   7.19 &  17.86 &  20.97 &  -24.07 \\ 
NGC3136                 &  151.4506 & -67.3780 &   15.3 &   0.16 &    -   &    -   &  RACS       &   7.22 &  24.32 &  20.90 &  -24.71 \\ 
NGC1574                 &   65.4951 & -56.9747 &   2.32 &   0.14 &    -   &    -   &  RACS       &   7.23 &  19.32 &  19.88 &  -24.20 \\ 
NGC4696                 &  192.2053 & -41.3111 &   3713 &   1.16 &   5591 &    -   &  RACS       &   7.26 &  35.48 &  23.79 &  -25.49 \\ 
NGC 3557                &  167.4905 & -37.5393 &    340 &   0.31 &    660 &   1313 &  Parkes     &   7.28 &  40.74 &  23.28 &  -25.77 \\ 
NGC7049                 &  319.7510 & -48.5620 &   47.3 &   0.20 &   66.8 &    -   &  RACS       &   7.32 &  29.92 &  21.72 &  -25.06 \\ 
NGC 7507                &  348.0316 & -28.5396 &   0.46 &   0.20 &    -   &    -   &  RACS       &   7.33 &  24.55 &  19.39 &  -24.62 \\ 
NGC 5061                &  199.5211 & -26.8372 &   0.79 &   0.19 &    -   &    -   &  RACS       &   7.34 &  23.77 &  19.59 &  -24.54 \\ 
\hline
\end{tabular}
\end{table*}

\begin{figure}
\begin{center}
\includegraphics[width=\columnwidth]{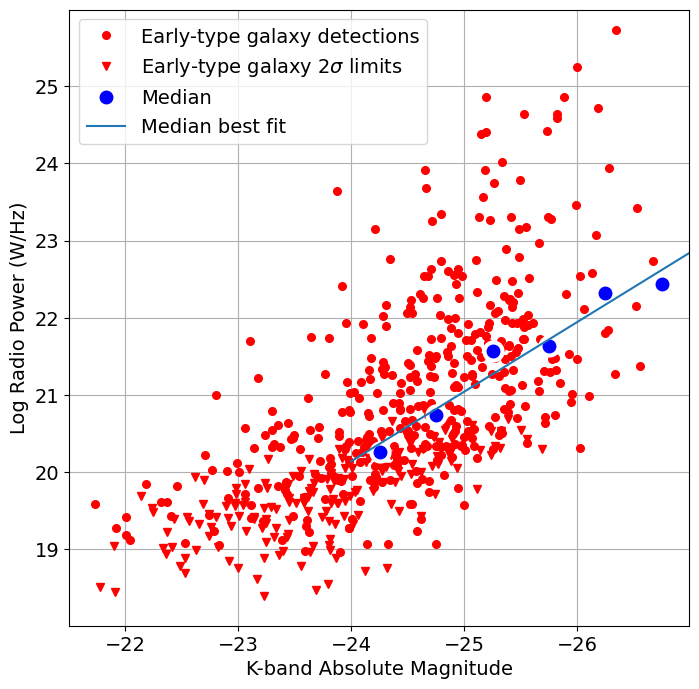}
\caption{Log 1.4~GHz power of $K_S<9.5$ early-type galaxies as a function of $K_S$-band absolute magnitude. For galaxies with a measured flux density per beam less than 2$\sigma$ above the local noise, we plot 2$\sigma$ upper limits, as shown with triangles. Large blue circles show median radio power as a function of absolute magnitude, which can be approximated by $P_{1.4} \propto L_K^{2.2}$. The distribution of log radio power is skewed, with some early-type galaxies having radio powers $\sim 10^3$ times higher than the median. All very massive early-type galaxies are radio sources, although some have radio powers comparable to that of the Milky Way.}\label{fig:lnu}
\end{center}
\end{figure}

Figure~\ref{fig:lnu} shows log radio power as a function of absolute magnitude for $K_S<9.5$ early-type galaxies. The distribution of radio powers is broad, and at fixed $M_K$ radio powers can span five orders of magnitude. While it appears the distribution of radio powers could narrow for $M_K>-24$ early-type galaxies, this may be an artifact of a decreasing fraction of early-type galaxies with radio detections. The distribution of log radio power is asymmetric, with some galaxies having radio powers $\sim 10^3$ times higher than the median. While \citet{brown2011} were able to fit a log-normal model to the distribution of radio powers (and limits) measured with NVSS as a function of absolute magnitude, such a model would be a poor fit to the higher quality RACS measurements.


Radio power is clearly a function of absolute magnitude in Figure~\ref{fig:lnu}, and we have determined the median radio power for absolute magnitude bins where 70\% of galaxies have RACS detections. A fit to the median radio power gives:
\begin{equation}
{\rm log} P_{1.4} = (21.49 \pm 0.08) - (0.90\pm 0.10) (M_K + 25.5),
\end{equation} 
which corresponds to $P_{1.4} \propto L_K^{2.2}$. This fit is shallower than our previous work, where we found $P_{1.4} \propto L_K^{2.8}$ \citep{brown2011}, which may be a consequence of using a (now deprecated) log-normal model. However, our fit is comparable to that of \citet{sadler1989}, who found $P_{30} \propto L_B^{2.2}$, where $P_{30}$ is the upper 30\% percentile of the 5~GHz radio power and $L_B$ is the $B$-band luminosity (although caution is required given the different percentiles, frequencies and optical-infrared bands used). 


\begin{figure}
\begin{center}
\includegraphics[width=\columnwidth]{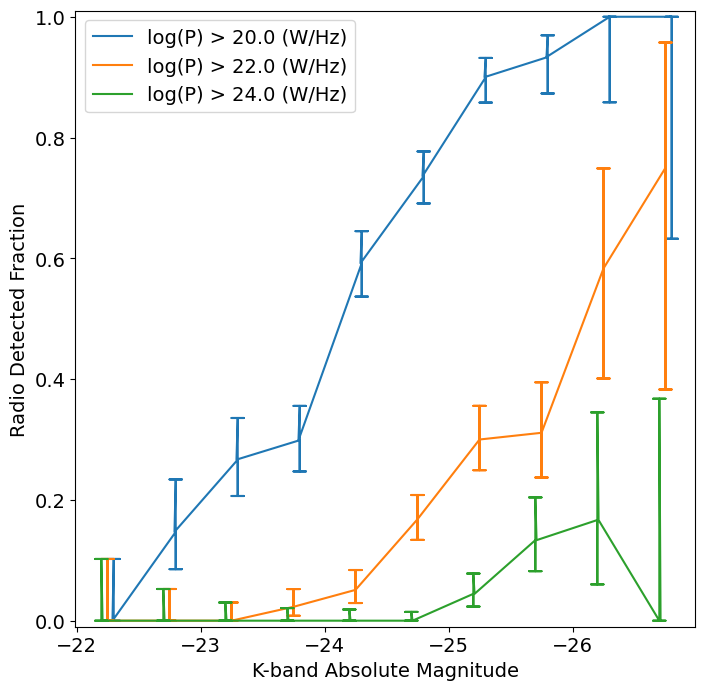}
\caption{The fraction of early-type galaxies above radio power thresholds as a function of $K$-band absolute magnitude. Binomial statistics have been used to determine the uncertainties and for clarity, the data for the different power thresholds has been slightly offset. 
\label{fig:frac}}
\end{center}
\end{figure}


In Figure~\ref{fig:frac} we present the fraction of early-type galaxies that are above radio power thresholds as a function of $K$-band absolute magnitude. Our combination of depth and low redshift is particularly useful for the statistics of the lowest power radio sources ($10^{20}~{\rm W/Hz}$). Binomial statistics indicate that at least 88\% of the overall $M_K<-26$ early-type population harbours a $10^{20}~{\rm W/Hz}$ radio source (95\% confidence), although we reiterate that 100\% of the most massive galaxies in our sample are detected with RACS.

\section{DISCUSSION}
\label{sec:discussion}


We have measured the radio powers of early-type galaxies with RACS, and found that all very massive early-type galaxies are radio sources. This builds on previous studies that found a large fraction of very massive early-type galaxies, brightest cluster galaxies (BCGs) and passive galaxies (without detectable star formation) are radio sources \citep{best2005,best2007,brown2011,sabater2019, capetti2022}, with \citet{brown2011}, \citet{sabater2019} and \citet{capetti2022} concluding all very massive galaxies are probably radio sources. 

\begin{figure}
\begin{center}
\includegraphics[width=\columnwidth]{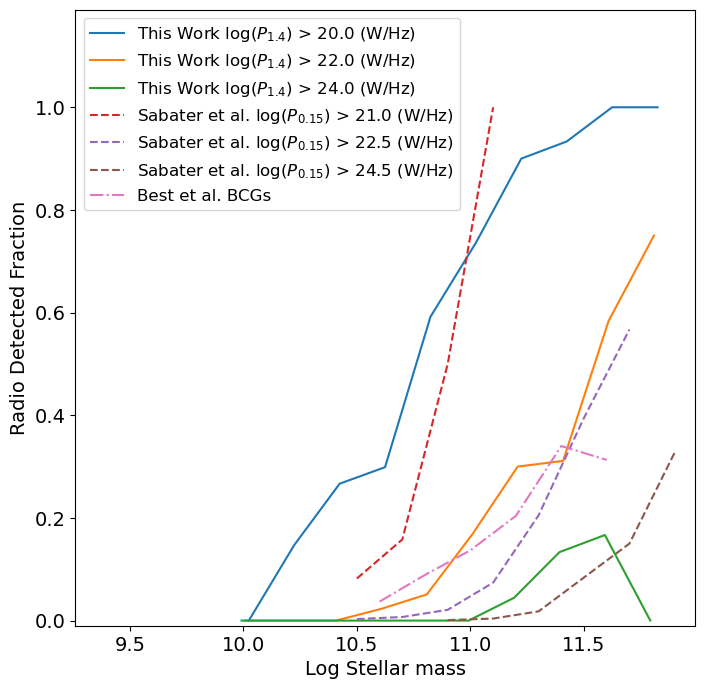}
\caption{The fraction of early-type galaxies that are radio sources as a function of stellar mass for this work, and the studies of \citet{best2007} for BCGs and \citet{sabater2019} passive galaxies. For clarity we do not show uncertainties, which can be large for the highest mass bins. While the different works have different galaxy selection criteria and observed at radio frequencies between 150~MHz and 1.4~GHz, the broad overall trends are comparable. 
\label{fig:fracmass}}
\end{center}
\end{figure}

In Figure~\ref{fig:fracmass} we compare our work with that of \citet{best2007} for BCGs and \citet{sabater2019} for passive galaxies, by plotting the fraction of galaxies that are radio sources as a function of stellar mass. To approximate stellar masses we have assumed a $K$-band mass to light ratio of $10^{-0.2}$ for passive galaxies, which is comparable to a \citet{kroupa2002} initial mass function. For clarity we have not shown error bars, and caution that for the highest mass bins the sample sizes can be small and the uncertainties large. For example, the two highest mass bins for the ${\rm log} P_{0.15}>21$ sample of \citet{sabater2019} contain only 5 galaxies. That said, for those bins where there are relatively large numbers of galaxies, we can see broadly similar trends, particularly for radio powers of $P \sim 10^{22}~{\rm W/Hz}$. 

Why can the radio powers of elliptical radio galaxies at fixed stellar mass differ by orders of magnitude? A simple solution would be environment, where powerful sources reside at the centres of relatively high mass dark matter halos with higher rates of gas in-fall. Alternatively, the observed distribution of radio powers could reflect a duty cycle with no dependence on environment, where radio powers of individual galaxies vary widely with time. In this latter scenario, the environments of high and low power radio sources would be comparable, as the large-scale environments of galaxies change relatively slowly with time. 

\begin{figure}
\begin{center}
\includegraphics[width=\columnwidth]{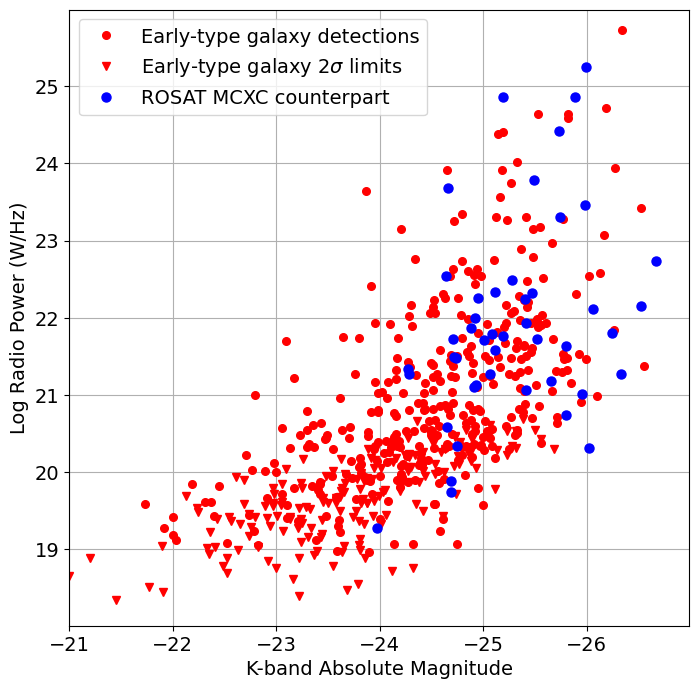}
\caption{Log 1.4~GHz power of $K_S<9.5$ early-type galaxies as a function of $K_S$-band absolute magnitude, with matches to the {\it ROSAT} MCXC highlighted. X-ray luminosity is strongly correlated with $K$-band luminosity (and stellar mass), but galaxies with both high and low radio powers have counterparts in the MCXC. \label{fig:rosat}}
\end{center}
\end{figure}

To briefly explore the correlation between radio power and environment, we have cross matched our elliptical galaxy sample with the {\it ROSAT} Meta-Catalogue of X-ray Detected Clusters of Galaxies \citep[MCXC; ][]{piffaretti2011}, where X-ray luminosity serves as a proxy for halo mass. As Figure~\ref{fig:rosat} illustrates, the highest mass galaxies often have X-ray counterparts as they reside within the most massive dark matter halos \citep[e.g. ][]{lin2004,brown2008}, but once this confounding factor is accounted for there is no obvious correlation between radio-power and X-ray counterparts in MCXC. The literature on radio galaxy environments reaches a range of conclusions, with some seeing modest correlations between radio power and environment once the confounding factor of host galaxy mass is accounted for \citep[e.g.][]{worpel2013}, while others see powerful radio sources preferentially in over-dense environments \citep[e.g.][]{mandelbaum2009}. Larger samples of low redshift radio galaxies will be selected with the completed ASKAP Evolutionary Map on the Universe \citep[EMU; ][]{norris2011} and should better address this question.

\begin{figure}
\begin{center}
\includegraphics[width=\columnwidth]{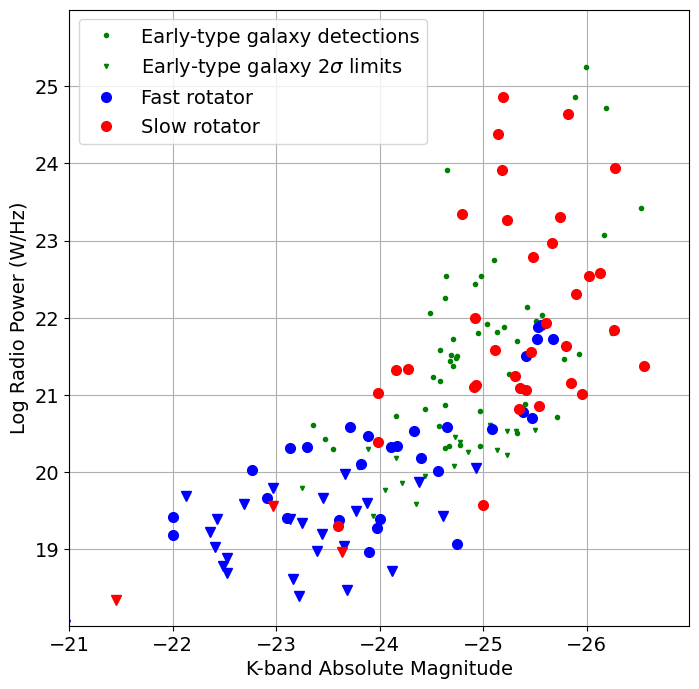}
\caption{Log 1.4~GHz power of $K_S<9.5$ early-type galaxies in the declination range $-6^\circ < \delta < +30^\circ$ as a function of $K_S$-band absolute magnitude, with galaxies with fast and slow rotators highlighted. While there is an obvious trend with $K$-band absolute magnitude, there's also a trend with radio power, and all early-type galaxies with radio powers above $10^{22}~{\rm W/Hz}$ are slow rotators. \label{fig:fastslow}}
\end{center}
\end{figure}

There are well-established correlations between AGN radio power and host galaxy properties, including stellar mass (as discussed earlier) and morphology (i.e. early-type galaxies). Within early-type galaxies there is a spread in morphological and kinematic properties, including fast and slow rotators that show relatively high and low net angular momentum compared to the observed dispersion \citep[e.g. ][]{kormendy1996,emsellem2007}. Within elliptical galaxies, \citet{bender1987,bender1989} found that galaxies with boxy isophotes typically had higher radio powers than galaxies with disk components. \citet{zheng2023}, using LoTSS and Very Large Array (VLA) radio surveys along with MANGA integral field spectroscopy, recently found that high power radio sources at fixed stellar mass are preferentially in early-type galaxies classified as slow rotators. North of $\delta = -6^\circ$ our sample overlaps with local integral field spectroscopy of early-type galaxies that provide fast and slow rotator classifications \citep{emsellem2007,emsellem2011,veale2017} using the $\lambda_{R_e}=0.25$ criterion where 
\begin{equation}
\lambda_{R} = 
\frac{\sum^{N_p}_{i=1} F_i R_i |V_i| }{\sum^{N_p}_{i=1} F_i R_i \sqrt{V_i^2+\sigma_i^2} }
\end{equation}
where $R_e$ is the effective radius, $R$ is the radius from the galaxy centre, $F$ is the surface brightness, $V$ is the stellar rotation velocity and $\sigma$ is the stellar velocity dispersion. In Figure~\ref{fig:fastslow} we show the $K$-band absolute magnitudes and radio powers of the $\delta > -6^\circ$ subset with fast and slow rotator classifications highlighted. 

In Figure~\ref{fig:fastslow} early-type galaxy angular momentum is clearly correlated with absolute magnitude (and stellar mass), but at the highest masses there is also a correlation between radio power and angular momentum, with high power radio sources being preferentially hosted by slow rotators. The distribution of radio powers at fixed absolute magnitude (or stellar mass) should thus not be considered a direct measurement of a duty cycle, as the host galaxies of high and low power radio sources are (on average) different. Given radio power in passive galaxies (at fixed stellar mass) correlates with both rotation and morphology, it is plausible that rotational support of gas results in fast rotators having (on average) lower accretion rates than slow rotators. Our trend is qualitatively similar to that observed by \citet{zheng2023} but they find some high power radio sources hosted by fast rotators whereas in our (admittedly) incomplete sample all the early-type galaxies hosting radio sources above $10^{22}~{\rm W/Hz}$ are slow rotators.

\begin{figure}
\begin{center}
\includegraphics[width=\columnwidth]{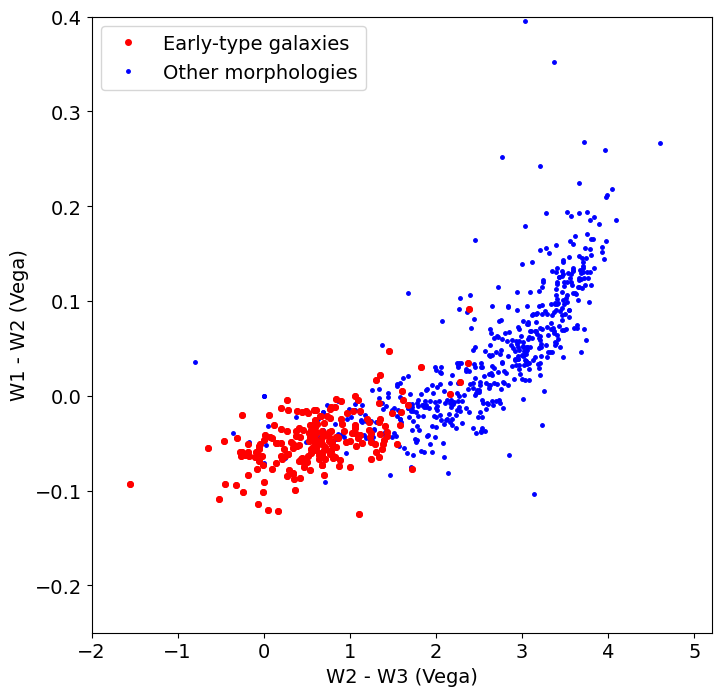}
\caption{The {\it WISE} colours of $K_S<9.5$ galaxies with RACS detections and radio powers below $10^{21}~{\rm W/Hz}$. As expected, most galaxies with early-type morphologies are near the origin of the diagram, corresponding to (approximately) the Rayleigh-Jeans spectral energy distributions of stellar populations and little infrared emission by warm dust. While the {\it WISE} colours of early-type galaxies are consistent with their radio sources being powered by AGNs, we cannot rule out very low levels of star formation in these galaxies. \label{fig:wise}}
\end{center}
\end{figure}

While AGNs clearly power $>10^{23}~{\rm W/Hz}$ radio sources in early-type galaxies, do they also power $< 10^{21}~{\rm W/Hz}$ radio sources in early-type galaxies? In Figure~\ref{fig:wise} we plot the Wide-field Infrared Survey Explorer \citep[{\it WISE};][]{wright2010, parkash2018, jarrett2019} colours of $K_S<9.5$ galaxies with radio powers below $10^{21}~{\rm W/Hz}$, with early-type galaxies highlighted in red. The {\it WISE} $W1$, $W2$ and $W3$ bands correspond to 3.4, 4.6 and $12~{\rm \mu m}$ respectively, with $12~{\rm \mu m}$ being sensitive to thermal emission from dust heated by star formation \citep[e.g.][]{wright2010}. The early-type galaxies primarily reside near the origin of the {\it WISE} colour-colour diagram, where the colours are comparable to a Rayleigh-Jeans spectra of stars with little emission from warm dust, consistent with passive galaxies (with little or no star formation) hosting AGNs.


While the {\it WISE} colours of our sample are consistent with them being powered by AGNs (i.e. not powered by star formation), this does not rule out other sources of radio emission. As noted earlier, \citet{capetti2022} conclude that for their 25 $M_K<-25.8$ galaxies with radio detections, at least four are AGN remnants and at least one is powered primarily by star formation. Also, a radio power of $\sim 10^{21}~{\rm W/Hz}$ corresponds to a star formation rate of $\sim 0.5~{\rm M_\odot~yr^{-1}}$ \citep[e.g.][]{brown2017} and for a very massive elliptical galaxy this corresponds to a specific star formation rate of $\sim 2\times 10^{-3}~{\rm Gyr^{-1}}$. Such low specific star formation rates may not be picked up by many star formation rate tracers (including {\it WISE} infrared colours), so it is possible some weak radio sources in our sample are powered by star formation without a contribution from an AGN (or an AGN remnant). 


\section{CONCLUSION}
\label{sec:conclusion}


We have measured radio continuum emission from 587 $K_S<9.5$ morphological early-type galaxies using 2MRS sample and the RACS 887.5~MHz images of galaxies in the declination range $-84.1^\circ<\delta<+30.0^\circ$. With a typical RMS depth of $300~{\rm \mu Jy/beam}$ and a $\sim 25^{\prime\prime}$ beam, we are able to measure radio emission from 70\% (408) of the early-type galaxies in our sample with more then $2\sigma$ significance. We find both radio flux density and radio power are a strong function of $K$-band absolute magnitude. All 40 of the $M_K<-25.7$ early-type galaxies in our sample have measured flux densities that are more than $2\sigma$ above the noise.



At fixed $K$-band magnitude the distribution of radio powers spans five orders of magnitude and is asymmetric, with some galaxies having radio powers more than 1000 times larger than the median. Comparing our sample to integral field spectroscopy kinematics from the literature, we find that high power radio sources are preferentially found in host galaxies with relatively low angular momentum (slow rotators). We thus conclude that the broad distribution of radio powers cannot be directly interpreted as a duty cycle, as (on average) the host galaxies of high and low power radio sources are different.


We conclude that all very massive ($>3 \times 10~{\rm M_\odot}$) early-type galaxies harbour radio sources, although in some instances the 1.4~GHz radio power is only $\sim 3\times 10^{20}~{\rm W~Hz^{-1}}$. The infrared colours of most early-type galaxies detected by RACS are comparable to passive galaxies without star formation, and for powerful radio sources this implies they are powered by AGNs. However, for the weakest radio sources it is plausible that very very low levels of star formation could produce low power radio sources with little impact on infrared colours. 

\begin{acknowledgements}

The Australian Square Kilometre Array Pathfinder is part of the Australia Telescope National Facility which is managed by CSIRO. Operation of ASKAP is funded by the Australian Government with support from the National Collaborative Research Infrastructure Strategy. ASKAP uses the resources of the Pawsey Supercomputing Centre. Establishment of ASKAP, the Murchison Radio-astronomy Observatory and the Pawsey Supercomputing Centre are initiatives of the Australian Government, with support from the Government of Western Australia and the Science and Industry Endowment Fund. We acknowledge the Wajarri Yamatji as the traditional owners of the Murchison Radio-astronomy Observatory site. This paper makes use of services or code that have been provided by CASDA and AAO Data Central. T. A. C. undertook preliminary work on this project as part of Monash University's PHS~3350 unit. T. A. C. receives support from the Australian Government Research Training Program. We thank an anonymous referee whose comments and suggestions significantly improved the manuscript. 

\end{acknowledgements}

\bibliographystyle{pasa-mnras}
\bibliography{racs_elliptical}

\end{document}